\documentstyle[epsfig]{l-aa}

\def\rerg{\rm erg}
\def\rs{\rm s}
\def\rs1{\rm s^{-1}}
\def\fxg{f_{X/\gamma} }

\def\rcm{\rm cm}
\def\rcm2{\rm cm^{-2}}
\def\deg{\rm ^{\circ}}
\def\flux{\rerg\ \rcm2\ \rs1}

\def\apj{{\it Astrophys. J. }}
\def\apjs{{\it Astrophys. J. Suppl. Ser. }}
\def\pasj{{\it Publs. Astron. Soc. Japan }}
\def\nature{{\it Nature }}
\def\aas{{\it Astron. Astrophys. Suppl. Ser.}}
\def\mnras{{\it Mon. Not. R. Astr. Soc.}}

\def\jref#1 #2 #3 #4 {{\par\noindent \hangindent=2em \hangafter=1
     \advance \rightskip by 0em #1, {\it#2}, {\bf#3}, #4.\par}}
\def\rref#1{{\par\noindent \hangindent=2em \hangafter=1
     \advance \rightskip by 0em #1.\par}}

\begin{document}
\thesaurus{011
           (13.07.1;   
            13.25.1;)} 

\title{The first X--ray localization of a
$\gamma$--ray burst by BeppoSAX and its fast spectral evolution }
\author{L. Piro\inst{1} \and J. Heise\inst{2} \and R. Jager\inst{2} \and 
E. Costa\inst{1} \and F. Frontera\inst{3} \and M. Feroci\inst{1} \and 
J. M. Muller\inst{2,5} \and L. Amati\inst{1,6} \and
M. N. Cinti\inst{1} \and D. Dal Fiume\inst{4} \and L. Nicastro\inst{4} \and 
M. Orlandini\inst{4} \and G. Pizzichini\inst{4}}

\institute{{Istituto Astrofisica Spaziale, C.N.R., Via E. Fermi 21, 00044 
Frascati, Italy}
\and
{Space Research Organization in the Netherlands, 
Sorbonnelaan 2, 3584 CA Utrecht, The Netherlands}
\and
{Dip. Fisica, Universita` Ferrara, Via Paradiso 12, Ferrara, Italy}
\and
{Istituto TeSRE, C.N.R., Via Gobetti 101, 40129 Bologna, 
Italy}
\and
{Beppo-SAX Scientific Data Center, Via Corcolle 19, 00131 
Roma, Italy}
\and
{Istituto astronomico, Universita` "La Sapienza", Via Lancisi 29, 
Roma, Italy}}

   \offprints{L. Piro:piro@alpha1.ias.fra.cnr.it}
   \date{Received 20 June 1997 / Accepted 15 July 1997}
  \maketitle
 \markboth{L.Piro et al.}{Precise X--ray localization }

\begin{abstract}
In this paper we present the observations performed by the BeppoSAX
Gamma--Ray Burst Monitor (GRBM) and Wide Field Cameras (WFC) of GB960720. 
We derive a precise localization (3 arcmin radius) and fast 
broad band ($2-700$ keV) spectral evolution  
of the event. A search in the catalogues at all wavelengths
in the error box 
yields a unique outstanding source: the bright radio quasar
4C 49.29. Although the probability of finding such a source 
by chance
is very low (  $\sim 2 \times 10^{-4}$), the absence of similar counterparts in
other small error boxes suggests a chance occurrence. 
We also find that the duration-energy relationship for bursts previously 
observed above 25
keV (Fenimore et al. 1995) extends down to 1.5 keV.
This result suggests that the same radiation mechanism is ope-rating from X--rays
to gamma--rays and is in agreement with radiative cooling by synchotron
emission.
A fast evolution of the spectrum is found, in which the ratio of X- to
gamma--ray intensities varies over three orders of magni-tude.
Furthermore, the spectrum in  the initial phase of the event betrays the 
presence
of an optically thick source rapidly evolving in a thin configuration.
No other class of sources in the universe shows such a fast and extreme 
evolution. 
These results pose new and tighter constraints on theoretical models for
gamma--ray bursts.

               \keywords{Gamma Ray Burst: localization -- spectral evolution}
   \end{abstract}

\section{Introduction }
More than twenty years after their discovery (Klebesadel et al. 1973), the 
origin of gamma--ray bursts 
is still one of the great mysteries in astrophysics. 
The main reason is the difficult identification of a counterpart.
Accurate position determination has been
provided in the past only for a few dozens events
(Atteia et al. 1987, Boer et al. 1994).
The  X--ray astronomy satellite BeppoSAX 
(Piro et al. 1995, Boella et al. 1997), a major 
program of the Italian
Space Agency (ASI) with participation of the Netherlands Agency for Aerospace
Programs (NIVR), represents a step forward, in that
it combines a gamma--ray burst monitor (GRBM) with
two X--ray wide field cameras (WFC),
allowing simultaneous detection,  3 arcmin localization 
and X to $\gamma$--ray spectral measurements
of GRB's.
The GRBM 
($50-700$ keV, Costa et al. 1997a) is based on four 
CsI scintillator slabs, orthogonal to the NFI and each one to the next one.
The WFC
(Jager et al. 1997) are coded mask proportional counters, 
operating in the $1.5-26$ 
keV
range. They  are pointed in opposite directions at 90$\deg$ from the axis of
Narrow Field Instruments (NFI) and are co-aligned with two of the GRBM slabs. 
They watch the X--ray sky
with a full field of view of $40\deg \times 40\deg$, 
with a spatial resolution of 5 arcmin (FWHM).
Positions can be derived with accuracies from 1 to 5 arcmin.

The paper is structured as follows: in sect. 2 we present the observation of
the event, discussing in particular the localization and the association with
a potential counterpart. In sect.3 we present the broad band 
light curves and discuss the relationship between the pulse duration and
energy. In sect. 4 we show the time resolved broad band spectra of the event
and discuss some of the implications derived from the
observed  strong spectral evolution of the event.
  
\section{Observation }

On July 20, 1996 11:36:53 UT a gamma--ray burst (GB960720) was simultaneously
detected in the GRBM and one WFC. The same event was detected by BATSE,
(trigger n.5545, BATSE team and BACODINE, private communication). 
The burst had a peak flux of 1300 cts/s in the GRBM
(corresponding to $F_{\gamma}\simeq 10^{-6} \flux$), and 82 cts/s
(corresponding to $F_X\simeq 2.5\times 10^{-8} \flux$ after correction for the
off-axis position) in the WFC. The total flux emitted was $2.5\times 10^{-6}
\rerg\ \rcm2$. The value of $\fxg \simeq 0.03$ of the peak luminosities of this
relatively weak burst is similar to that found in brighter events
(Laros et al. 1982, Yoshida et al. 1989).

The sky image of the full field of view of the WFC obtained in 15 s on the
burst is shown in Fig. 1. a)

\begin{figure}
\label{fig1ab}
\epsfig{figure=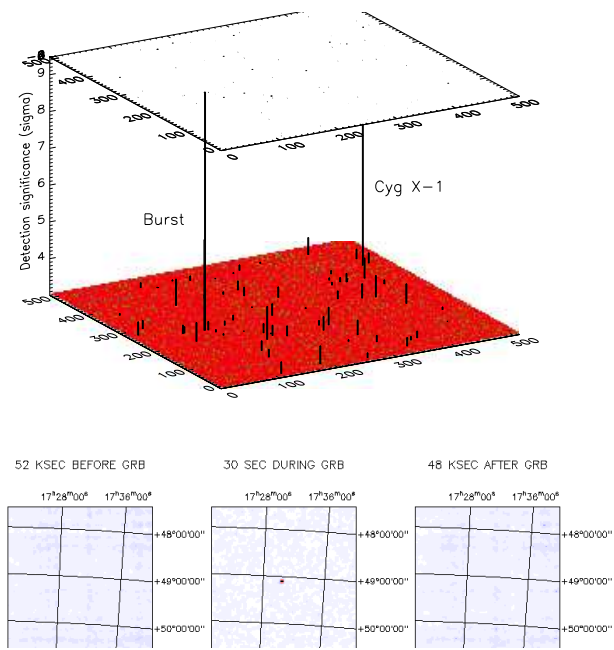, width=12cm,
 bbllx=120pt , bblly=340pt, bburx=660pt, bbury=720pt,clip=yes}
\caption{
{\bf a)} The $40\deg \times 40\deg$ image in the $2-26$ keV range 
of the WFC integrated on a period
of 15 s on the burst. The y axis gives the significance of the detection
($\sigma$). The source close to the edge is Cyg X-1.
{\bf b)} Images 
of the field centered on GB960720 in time sequence.
The first and last images were obtained  integrating
over $\sim\ 50,000 s$ before and after the burst.
 No source was detected.
The second image is a 30 s long shot which shows the 
sudden presence of  GB960720. 
We find a 90\% confidence level error circle for GB960720 that is centred on $\alpha$(2000)=17h30m36s, $\delta$(2000)=+49$\deg$ 05' 49" and has a radius
of 3 arcmin.
The position uncertainties have a statistical  component 
depending on the source strength and a systematic component 
depending on the offset position in the field of view. Independent 
checks of the images scales, and a modelling of the spectrally and 
positionally dependent point spread function have been performed.
The distribution of the error was checked with 93 independent sources.}
\end{figure}

 The 40$\deg$ wide field shows the burst and, close to
the edge, the well known X--ray source Cyg X-1. The Palomar Survey shows several
objects within the 3 arcmin radius error circle. The only one 
identified in catalogues is 
the radio loud quasar 4C 49.29, with a flux of  2
Jy at 408 MHz (Reid et al. 1995). 
Considering all the
radio sources with a flux brighter than that of the quasar 
(Colla et al. 1972) and the fraction of quasars in the radio 
source population at this flux level (Singal et al. 1993) the probability
of chance occurence is about {$2 \times 10^{-4}$}. 
The association of the gamma ray burst with
a radio loud quasar is very tempting,  considering the environment and
radiation mechanisms supposedly operating in these two classes of sources. 
One of the most common explanation of the gamma--ray burst phenomenon, the
fireball model (Rees \& M\'esz\'aros 1992), involves relativistic expansion of a population of
electrons that radiates mostly via synchrotron process. This is
similar, but for the much greater Lorentz factor involved in the fireball
models, to the scenario of radio loud quasars.
However this is the only radio loud quasar  
found in a small gamma--ray burst error box, so that it may
be a chance occurence. More data are
needed to confirm the association.
 
A deep pointing (56 ksec)
of the field was performed on Sept. 3, 1996
 with the Narrow Field Instruments. The 45 days delay between burst             
trigger and NFI observation of the WFC error box is due to 
the fact that the simultaneous detection in WFC and GRBM of the event
 was discovered
during off-line analysis.
The NFI data confirm that a faint X-ray emission (F(2-10
keV)=$(1.0\pm0.3)\times 10^{-13} \flux$) is associated with the
quasar (Piro et al. 1996).  However both the X-ray luminosity
 of $4\times10^{44}$ {\rm
erg \ s$^{-1}$ (at a redshift z=1.038 (Reid et al. 1995), H$_{0}$=50 km/s/Mpc) 
and the ratio of the X--ray to the
optical flux ($m_V=18.8$ ( Walsh et al. 1984)) $f_X/f_V=1$ are typical
of a quasar  (Maccacaro et al. 1980) and therefore not immediately related to afterglow activity of the
gamma--ray burst.

Information on afterglow activity in X--rays on much shorter time scales can be
derived from the observation with the WFC. The observation covers a period of
about 100,000 s centered in time on the event. In Fig.1b we show the images of
the field in time sequence. No si-gnificant emission is
detected above the background either before of after the burst,
 with a $5\sigma$ upper limit of $\sim 10^{-10}
\flux$, 200 times lower than the peak flux. Any signature of an X--ray halo
produced by dust scatte-ring in our Galaxy, --- whose measurement would provide 
a direct test for a local vs. galactic halo / extragalactic origin
(Pacinski 1991, Klose 1994) --- would remain below the upper limit of the WFC by
three orders of magnitude.
\section { Pulse duration from X--rays to gamma--rays}
Simultaneous X--ray to gamma--rays observation provide important information to
constraint radiative mechanisms operating in gamma--ray bursts.
The light
curve in Fig. 2 shows that the burst duration increases with decreasing 
energy, from about 2 s in the $100-700$ keV range to about 16 s in 
the $1.5-3.5$
keV range. This behaviour was observed in the past by other spacecraft 
experiments
(Laros et al. 1982, Yoshida et al. 1989).

\begin{figure}
\label{fig2}
\epsfig{figure=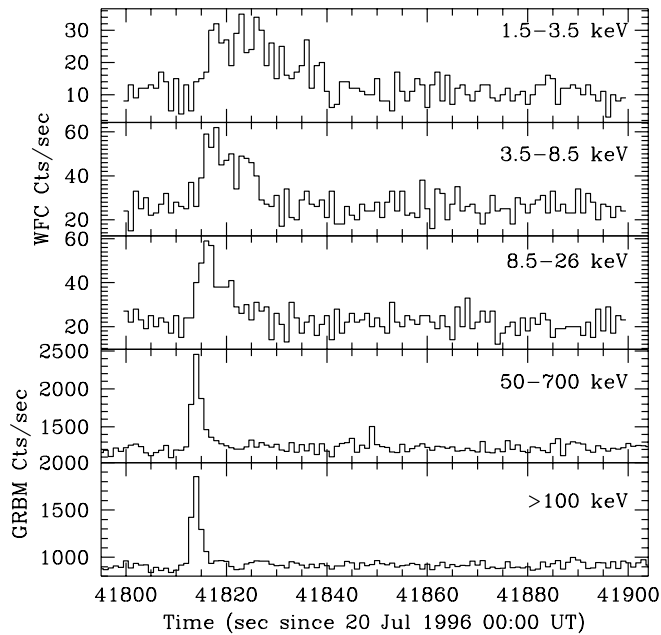,height=9.0cm,width=9cm, angle=0}
\caption{
Light curves, by raw data, of the burst in four energy ranges: $2-8$ keV,
$8-26$ keV (WFC); $50-700$ keV, above 100 keV (GRBM).}
\end{figure}
 
In Fig. 3 we show that this dependency can be expressed as

\begin{equation}
\Delta t \sim E^{-0.46}
\end{equation}

\noindent
in the whole band from 1.5 to 700 keV. 
This is consistent with
 BATSE results at higher energies (Fenimore et al. 1995),
suggesting that the emission mechanism is the same from 
soft X--rays to gamma--rays. In particular it is consistent
with what predicted by
radiative cooling by synchrotron losses (Tavani 1996). On the reverse 
eq. (1) argues against a scenario in which X--rays and gamma--rays are 
produced by the same population of 
electrons via synchrotron and Inverse Compton processes respectively (Self--Synchro--Compton). In this case we would
expect to observe a similar duration in the X--ray and gamma--ray
range. In fact, an electron with Lorentz factor $\gamma$ will produce X--ray and
gamma--ray photons with the same cooling time $\tau=(1/\tau_S+1/\tau_{IC})^{-1}=
{\rm min} [\tau_S,\tau_{IC}]$. Eq. (1) also argues against models where the
duration of the burst is independent of energy, as in some realizations of the 
fireball models, where the burst duration is determined by the hydrodynamical
time scale (Sari et al 1996).

\begin{figure}
\label{fig3}
\epsfig{figure=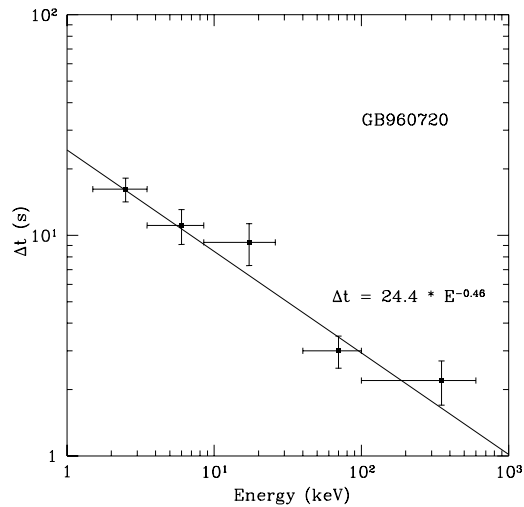,height=7.0 
cm,width=7 cm, angle=0}
\caption{
The relationship $\Delta t$ vs E for GB960720. $\Delta t$ is the 
time taken to go from 5\% to 85\% of the total counts. 
The best fit gives an index of 0.46$\pm0.10$.}
\end{figure}   

\section{Broad band  spectral evolution}
Hard to soft spectral evolution appears to be a common -- though not ubiquitous -- property of gamma--ray
bursts (Yoshida et al. 1989);
this is also a distinct feature of this burst, as clear From Fig. 2.
 Moreover, from Fig. 4, a blow up of the
initial part of the GRB, it is evident how 
the rise time remarkably depends on photon energy.
The event starts with a sudden emission of gamma--ray photons 
(Fig. 4, lower panel), while the emission
at lower energies is negligible (Fig. 4, upper panel).
\begin{figure}
\label{fig4}
\epsfig{figure=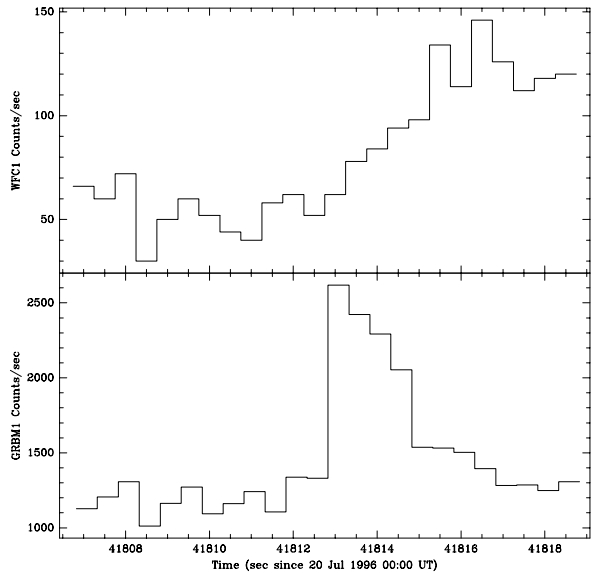,height=8.5cm,width=8.5cm, angle=0}
\caption{
Light curves with 0.5 s bin of GB960720 in the
$2-26$ keV range (upper panel) and $50-700$ keV range (lower panel).}
\end{figure}
As the
gamma--ray emission decays with a time scale of $\sim\ 3.5$ s, the X--ray
luminosity increases with a similar time scale. This behaviour can be better
 understood by analyzing the time resolved spectra
(Fig. 5) which show indeed a strong variation, with the power law photon
index ($N\sim E^{\alpha}$) varying from $\alpha>0.1$ to
$\alpha=-2.4\pm0.7$. After a few seconds the radiative output channel of the
source goes from photons with E$ \sim$ $100-700$ keV ($\fxg< 3\times 10^{-3}$, as
from Fig. 5), to photons with E $\sim$ $1.5-3.5$ keV ($\fxg>1$).  

\begin{figure}
\label{fig5}
\epsfig{figure=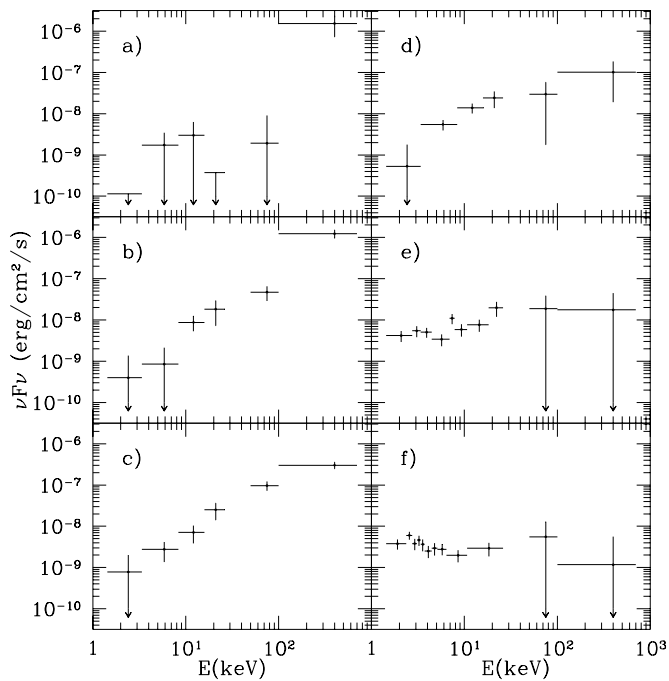,height=9.0cm,width=9cm, angle=0}
\caption{
Spectral evolution of GB960720. The different pannels correspond to 
the following time intervals from burst on--set (see also first column of 
Table 1. ):
${\bf a)}$ from 0 to 1 second,  
${\bf b)}$ from 1 to 2 second, 
${\bf c)}$ from 2 to 3 second, 
${\bf d)}$ from 3 to 4 second, 
${\bf e)}$ from 4 to 8 second, 
${\bf f)}$ from 8 to 17 second }
\end{figure}

This result
extends to lower energies the spectral evolution observed above $\sim\ 25$ keV
by BATSE, which shows that the peak of the energy distribution decreases
substantially during the burst decay (Ford et al. 1995, Liang \& Kargatis 1996).

We note that in the first second of the event there is an 
indication that the spectral photon index is higher than
the maximum value ($-0.67$) expected in the case of 
optically thin synchrotron emission (Ribicki \& Lightman 1979).
A similar evidence at the higher energies covered by BATSE has been observed 
by Crider et al. (1997).
If synchrotron 
emission
is the mechanism producing the photons of the burst, then our result could
be explained with
the presence of self-absorption in the initial part of the event. The source
then quickly evolves and after 2 seconds the spectral index becomes consistent
with that expected in the optically thin case (table 1). 

Although the data do not permit to constrain complex spectral shapes,
we have attempted to verify the consi-stency of the data with the scenario
expected from synchrotron emission.
We have assumed a fast evolving synchrotron spectrum 
characterized by  self
absorption at $E_c$ and a break frequency at $E_0$ that should
correspond to
the minimum energy of the electron distribution.
Below $E_c$ we have  assumed the optically thick 
slope of 1.5, while
between $E_c$ and $E_0$ we have adopted  a slope of $-0.67$,
e.g. that expected by  optically thin
synchrotron emission below the minimum energy of the electron distribution.
Above $E_0$  the  emission is determined by the electron distribution,
and in this case we have adopted a power law with slope fixed to 
the value of 2.4, 
the asymptotical value observed in the last
part of the event. The connection between the different power laws has
been modelled via a Band formula (Band et al., 1993). 
In order to further reduce the number of free parameters, we have assumed
that in the first 2 seconds $E_0$ is at energies greater than
700 keV, and that in the following  evolution $E_c$ has an energy below
1 keV.
The only free parameters were then one of the two
 break energies and the normalization.
From Table 1 (right colums) we see that in the first 2 seconds $E_c$
evolves very rapidly from hard X--rays to a few keV.
In the following evolution a similar behaviour is observed for $E_0$.  

As mentioned before, this spectral evolution could
explain well
the shape of the light curves observed at the different  energies.
The X-ray emission in the first seconds corresponds
to a photon spectrum rising in energy, hence  it increases as
this spectrum shifts to lower energies.
In a similar fashion, the shift 
of the steep part of the spectrum produces the
fast decrease in 
hard-X-rays. In this picture, the observed X-rays peak delay is naturally explained
with the synchrotron self-absorption assumed above for the first $1-2$
seconds spectra.


\begin{table}

      \caption{Fits of time resolved spectra} 
         \label{powlaw}
\begin{tabular}{||c|cl|c||cl|c||}                                      \hline
$\Delta t$ & $\alpha  ^{\rm (a)}$   &       & $\chi^{2}_{\nu}$  & 
$E_{c}$ $ ^{\rm (b)}$, $E_{0}$ $ ^{\rm (c)}$        &    & 
$\chi^{2}_{\nu}$ 
                                  \\
(sec)   &   &      &  &(keV)& & 
                                   \\ \hline  
0--1       & ${> 0.1}$  &      &            & 
523$^{+50}_{-49}$&$^{\rm (b)}$   & 0.95   
                                    \\ \hline 
1--2       &-0.39$^{+0.23}_{-0.18}$& &1.15   & 
6.31$^{+4.2}_{-3.4}$  & $^{\rm (b)}$ & 1.42   
                                    \\ \hline 
2--3     &-0.92$^{+0.19}_{-0.15}$&  &1.22   & 
271.9$^{+122}_{-69}$  & $^{\rm (c)}$ &0.63  
                                    \\ \hline 
3--4     &-1.18$^{+0.20}_{-0.27}$&  &1.37   & 
50.1$^{+35}_{-21}$    & $^{\rm (c)}$  &0.68 
                                    \\ \hline 
4--8     &-1.64$^{+0.59}_{-0.47}$&  &1.09   & 
8.5$^{+8.2}_{-4.0}$   & $^{\rm (c)}$  &2.1 
                                    \\ \hline 
8--17     &-2.44$^{+0.62}_{-0.66}$&  &0.7  & 
${< 2}$               & $^{\rm (c)}$  &0.7 
                                    \\ 
\hline 
\end {tabular}

\begin{list}{}{}
\item[$^{\rm (a)}$] Fit with simple power law
\item[$^{\rm (b)}$] Fit with self--absorbed syncrothron model
\item[$^{\rm (c)}$] Fit with Band form
\end{list}
\end {table}

\smallskip

This result challenges some
realizations of fireball mo-dels that typically foresee self-absorption at much
lower energies (M\'esz\'aros et al. 1994), although it should be noted that 
those models
assume a time averaged situation.

\section{Conclusions}
The gamma ray bursts localization capabilities of BeppoSAX described and 
demonstrated in this paper have been fully exploited by implementing 
procedures for accurate real-time position determination of  
GRBM + WFC simultaneously detected events. This has made possible
the discoveries of X-ray afterglows (e.g. Costa et al. 1997b) and possible counterparts at optical
and radio wave lengths, providing also new  and unprecedented X-ray data.
These and future observations are needed to understand the general
properties of the X to gamma--ray spectral evolution in gamma--ray bursts.
Nevertheless the results presented here, 
that may represent the most extreme example
of a common property of these events (Laros et al. 1982, Yoshida et al. 1989),
demonstrate the importance of wide-band spectral measurements of GRBs and 
demand more detailed theoretical efforts on spectral evolution to
account for the three order of magnitude variation of $\fxg$ and
its extremely small value in the first phase of the event.

\begin{acknowledgements} We thank the Beppo-SAX team for the efforts on the
mission, the BACODINE team for the distribution of burst coordinates, G. 
Fishman and M. McCollough of the BATSE team for the light curves of the event,
D. Frail for discussion on the radio quasar, M. Tavani and M. Vietri for
helpful discussions on some of the theoretical implications of 
this observation and G.C. Perola for his critical reading of a 
previous version of this paper.
\end{acknowledgements}

\end{document}